\documentclass[a4paper, 12pt, openany, oneside]{article}
\usepackage[cp1251]{inputenc} 
\usepackage[english, russian]{babel} 
\usepackage{amsmath, latexsym, amssymb, array, graphics,
amsfonts, amsmath, bm}

\makeatletter

\renewcommand{\Re}{\mathop{\rm Re\,}}
\renewcommand{\Im}{\mathop{\rm Im\,}}

\makeatother
\usepackage{indentfirst}

\makeatother
\renewcommand{\baselinestretch}{1.2}
\usepackage[dvips]{graphicx}
\usepackage[flushleft,small]{caption2}

\begin{document}

\thispagestyle{empty}
\large
\renewcommand{\abstractname}{\, Реферат}
\renewcommand{\refname}{\begin{center} REFERENCES\end{center}}

\begin{flushright}\it\Large
Dedicated to the Memory\\ of our Teachers\\ C. Cercignani and
K. Case
\end{flushright}

 \begin{center}
\bf Analytical solution of the second Stokes problem on behaviour of gas over a oscillation surface. Part I: eigenvalues and eigensolutions
\end{center}\medskip
\begin{center}
  \bf V. A. Akimova\footnote{$vikont\_ava@mail.ru$},
  A. V. Latyshev\footnote{$avlatyshev@mail.ru$} and
  A. A. Yushkanov\footnote{$yushkanov@inbox.ru$}
\end{center}\medskip

\begin{center}
{\it Faculty of Physics and Mathematics,\\ Moscow State Regional
University, 105005,\\ Moscow, Radio str., 10--A}
\end{center}\medskip

\tableofcontents
\setcounter{secnumdepth}{4}


\begin{abstract}
В настоящей работе сформулирована вторая задача Стокса о поведении разреженного газа, заполняющего полупространство. Плоскость,
ограничивающая полупространство, совершает гармонические колебания
в своей плоскости. Используется кинетическое уравнение с модельным интегралом столкновений в форме $\tau$--модели. Рассматривается случай диффузного отражения молекул газа от стенки.
Находятся собственные решения (непрерывные моды) исходного кинетического уравнения, отвечающие непрерывному спектру. Изучаются свойства дисперсионной функции. Исследуется дискретный спектр задачи, состоящий из нулей дисперсионной функции в комплексной плоскости. Показано, что число нулей дисперсионной функции равно удвоенному индексу коэффициента задачи. Под коэффициентом задачи понимается отношение граничных значений
дисперсионной функции сверху и снизу на действительной оси.
Далее находятся собственные решения (дискретные моды) исходного кинетического уравнения, отвечающие дискретному спектру.

В конце работы составляется общее решение кинетического уравнения в виде
разложения по собственным решениям с неизвестными коэффициентами, отвечающими дискретному и непрерывному спектрам.
\medskip


\medskip

transport theory of gases.
\end{abstract}

\begin{center}
\item{} \section{\bf Introduction}
\end{center}

Задача о поведении газа над движущейся поверхностью в последние годы
привлекает пристальное внимание \cite{Stokes} -- \cite{15}. Это связано с развитием современных технологий, в частности, технологий наноразмеров.
В \cite{Yakhot} -- \cite{15} эта задача решалась численными или
приближенными методами. В настоящей работе показано, что эта задача допускает аналитическое решение. Аналитическое решение строится с помощью теории обобщенных функций и сингулярных интегральных уравнений.

Впервые задача о поведении газа над стенкой, колеблющейся в своей плоскости, была рассмотрена Дж. Г. Стоксом \cite{Stokes}. Задача решалась гидродинамическим методом без учёта эффекта скольжения. Обычно такую задачу называют второй задачей Стокса \cite{Yakhot}--\cite{SS-2002}.

В последние годы на тему этой задачи появился ряд публикаций. В работе \cite{Yakhot} рассматривается бесконечная колеблющаяся поверхность. Задача рассматривается для любых частот колебания поверхности. Из кинетического уравнения БГК получено уравнение типа гидродинамического. Рассматриваются гидродинамические граничные условия. Вводится коэффициент, связывающий скорость газа на поверхности со скоростью поверхности. Изотермическое скольжение не учитывается. Получен вид графика зависимости силы трения на поверхности от частоты колебаний поверхности. Показано, что в случае высокочастотных колебаний сила трения, действующая на поверхность, не зависит от частоты.

В работе \cite{SK-2007} получены коэффициенты вязкостного и теплового скольжения с использованием различных модельных уравнений. Использованы как максвелловские граничные условия, так и граничные условия Черчиньяни --- Лэмпис \cite{Lampis}.

Наиболее близкая к решенной в первой и второй главах диссертации \cite{15} задача решена в статье \cite{10}: рассматривается газовый поток над бесконечной пластиной, совершающей гармонические колебания в собственной плоскости. Найдена скорость газа над поверхностью и сила, действующая на поверхность со стороны газа. Для случая низких частот задача решена на основе уравнения Навье --- Стокса. Изотермическое скольжение не учитывалось. Для произвольных скоростей колебаний поверхности задача решена численными методами на основе кинетического уравнения Больцмана с интегралом столкновений в форме БГК (Бхатнагар, Гросс, Крук). При этом рассматривался только случай чисто диффузного отражения молекул от поверхности. Дано аналитическое решение для случая колебаний высокой частоты. И в этом случае рассматривается только чисто диффузное отражение молекул от поверхности. В конце второй главы диссертации \cite{15} проведено сопоставление результатов, полученных в статье \cite{10} c результатами, полученными в диссертации \cite{15}.

Работа \cite{11} является экспериментальным исследованием. Изучается поток газа, создаваемый механическим резонатором при различных частотах колебания резонатора. Эксперименты показывают, что при низких частотах колебаний резонатора, действующая на него со стороны газа сила трения прямо пропорциональна частоте колебания резонатора. При высоких частотах колебания резонатора ($~10^8$ Гц) действующая на него сила трения от частоты колебаний не зависит.

В последнее время задача о колебаниях плоской поверхности в собственной плоскости изучается и для случая неньютоновских жидкостей \cite{5} и
\cite{6}.

В статье \cite{12} рассматривается пример практического применения колебательной системы, подобной рассматриваемой во второй задаче Стокса, в области нанотехнологий.

Общим существенным недостатком всех упомянутых теоретических работ по решению второй задачи Стокса является отсутствие учёта характера взаимодействия с поверхностью, т.е. рассматривается только случай полной аккомодации тангенциального импульса.

Коэффициент аккомодации тангенциального импульса является величиной, зависящей от состояния поверхности. И если в "естественном"\, состоянии значение этой величины как правило близко к единице, то при специальной обработке поверхности её значение можно уменьшить многократно \cite{13}, а значит и существенно изменить характер взаимодействия поверхности с прилегающим газом.

В условиях стремительного развития вакуумных технологий и нанотехнологий, совершенствования авиационной и космической техники весьма актуальным и целесообразным является развитие направления исследований, связанного с определением влияния взаимодействия молекул с поверхностью на перенос импульса в системе "газ -- твёрдое тело"\, при произвольном разрежении газа и установлением связи физических свойств межфазной границы с макроскопическими газодинамическими параметрами.

В диссертации \cite{15} были предложены два решения второй задачи Стокса, учитывающие весь возможный диапазон коэффициента аккомодации тангенциального импульса.
Эти решения отвечают соответственно гидродинамическому и кинетическому описанию поведения газа над колеблющейся поверхностью в режиме со скольжением.


В предлагаемой серии работ вторая задача Стокса впервые решается аналитически. При этом используются сингулярные интегральные уравнения с ядром Коши и обобщенные функции. Настоящая работа --- первая из этой серии.

В п. 3 настоящей работы рассматривается постановка второй задачи Стокса. Задача формулируется в общей постановке --- с использованием граничных условий Максвелла (зеркально -- диффузных граничных условий). Далее задача будет рассматриваться только для диффузных граничных условий. В качестве кинетического уравнения рассматривается линеаризованное кинетическое уравнение. Это уравнение получается путем линеаризации модельного кинетического уравнения Больцмана и интегралом столкновений в форме релаксационной $\tau$--модели. Это так называемое модельное кинетическое БГК (Бхатнагар, Гросс, Крук) уравнение.

Пластина (плоскость), ограничивающая полупространство с разреженным газом совершает колебательные движения вдоль оси $y$. В качестве граничных условий используются два условия. Одно из них --- граничное условие вдали от стенки --- требует исчезания функции $h(x_1,\mu)$ вдали от стенки. Второе условие --- условие на стенке --- вытекает из требования диффузного отражения молекул от стенки. Требуется определить функцию распределения газовых молекул, найти скорость газа в полупространстве и непосредственно у стенки, найти силу трения, действующую со стороны газа на пластину, найти мощность диссипации энергии пластины.

В п. 4 кинетическое уравнение упрощается путем представления функции распределения в виде произведения $y$--компоненты скорости молекул газа на новую неизвестную функцию. При этом получается однопараметрическое семейство кинетических уравнений с чисто мнимым параметром. Параметром уравнений служит безразмерная величина частоты колебаний пластины. Эта величина $\omega_1=\omega/\nu=\omega \tau$ равна частоте колебаний пластины $\omega$, деленной на величину частоты $\eta$ столкновений молекул газа, $\tau=1/\nu$ -- время между двумя последовательными столкновениями молекулы.

В п. 5 находятся собственные решения (непрерывные моды) исходного кинетического уравнения, отвечающие непрерывному спектру. Последний совпадает с действительной положительной полуосью. Эти собственные решения находятся в пространстве обобщенных функций (распределений). Приводятся формулы Сохоцкого для дисперсионной функции задачи, являющейся основной в построенной теории.
Дисперсионная функция является кусочно аналитической функцией с действительной осью в качестве линии скачков (разрывов).

В п. 6 исследуется дискретный спектр задачи, состоящий из нулей дисперсионной функции в комплексной плоскости. Применяется принцип аргумента из теории функций комплексного переменного. Показано, что число нулей дисперсионной функции равно удвоенному индексу коэффициента задачи. Под коэффициентом $G(\mu)$ задачи понимается отношение граничных значений
дисперсионной функции сверху и снизу на действительной оси:
$G(\mu)=\lambda^+(\mu)/\lambda^-(\mu)$. Выясняется, что существует критическая частота
$$
\omega_1^*=\max\limits_{0<\mu<+\infty}\sqrt{[\Im\lambda^+(\mu)]^2-
[\Re\lambda^+(\mu)]^2}\approx 0.733,
$$
такая, что при $\omega_1\in[0,\omega_1^*)$ индекс коэффициента задачи равен единице: $\varkappa(G)=1$, а при $\omega_1\in (\omega_1^*,+0)$ индекс коэффициента задачи равен нулю: $\varkappa(G)=0$. Таким образом, если $\omega_1$ находится первом (левом) регионе, то дисперсионная функция имеет два комплексно -- значных нуля, отличающихся лишь знаками в силу четности дисперсионной функции. Если параметр $\omega_1$ находится во втором (правом) регионе, то индекс задачи равен нулю, т.е. дисперсионная функция комплексно -- значных нулей не имеет.

Далее находятся собственные решения (дискретные моды) исходного кинетического уравнения, отвечающие дискретному спектру.

В конце работы составляется общее решение кинетического уравнения в виде суммы собственного дискретного решения, умноженного на неизвестную постоянную, и интеграла по непрерывному спектру от собственных решений, отвечающих непрерывному спектру, умноженных на неизвестную функцию.
Эти неизвестные постоянная и функция называются соответственно коэффициентами дискретного и непрерывного спектров. Эти неизвестные находятся из граничных условий. Этому вопросу будут посвящены следующие наши работы.

\begin{center}
 \item{}\section{Линеаризованное кинетическое уравнение для задачи о колебаниях газа}
\end{center}

Пусть разреженный одноатомный газ занимает полупространство $x>0$
над плоской твердой поверхностью, лежащей в плоскости $x=0$.
Поверхность $(y,z)$ совершает гармонические колебания вдоль оси $y$
по закону $u_s(t)=u_0\cos \omega t$.

Рассмотрим линеаризованное кинетическое уравнение
$$
\dfrac{\partial \varphi}{\partial t}+v_x\dfrac{\partial \varphi}{\partial x}+\varphi(t,x,\mathbf{v})=\dfrac{\nu m}{kT}v_yu_y(t,x).
\eqno{(1.1)}
$$

В (1.1) $\nu=1/\tau$ -- частота столкновений газовых молекул, $\tau$ -- время между двумя последовательными столкновениями молекул, $m$ -- масса молекулы, $k$ -- постоянная Больцмана, $T$ --
температура газа, $u_y(x)$ -- массовая скорость газа,
$$
u_y(t,x)=\dfrac{1}{n}\int f(t,x,\mathbf{v})d^3v,
\eqno{(1.2)}
$$
$n$ -- числовая плотность (концентрация) газа. Концентрация газа и его температура считаются постоянными в линеаризованной постановке задачи.

Введем безразмерные скорости и параметры: безразмерную скорость молекул:
$\mathbf{C}=\sqrt{\beta}\mathbf{v}$ \;$(\beta=m/(2kT))$, безразмерную массовую скорость $U_y(t,x)=\sqrt{\beta}u_y(t,x)$, безразмерное время $t_1=\nu t$ и
безразмерную скорость колебаний пластины $U_s(t)=U_0\cos\omega t$,
где $U_0=\sqrt{\beta}u_0$ -- безразмерная амплитуда скорости колебаний границы полупространства. Тогда уравнение (1.1) может быть записано в виде:
$$
\dfrac{\partial \varphi}{\partial t_1}+C_x\dfrac{\partial \varphi}{\partial x_1}+\varphi(t_1,x_1,\mathbf{C})={2C_y}U_y(t_1,x_1).
\eqno{(1.3)}
$$

Заметим, что для безразмерного времени $U_s(t_1)=U_0\cos\omega_1t_1$.

В задаче о колебаниях газа требуется найти функцию распределения $f(t_1,x_1,\mathbf{C})$ газовых молекул. Функция распределения свзана
с функцией $\varphi(t_1,x_1,C_x)$ соотношением:
$$
f(t_1,x_1,\mathbf{C})=f_M(C)\big[1+\varphi(t_1,x_1,C_x)\big],
\eqno{(1.4)}
$$
где
$$
f_M(C)=n\Big(\dfrac{\beta}{\pi}\Big)^{3/2}\exp(-C^2)
$$
-- есть абсолютный максвеллиан.

Затем на основании найденной функции распределения требуется найти массовую скорость газа, значение
массовой скорости газа непосредственно у стенки. Кроме того, требуется вычислить силу сопротивления газа, действующую на колеблющуюся пластину, ограничивающую газ.

Подчеркнем, что задача о колебаниях газа решается в линеаризованной постановке.
Линеаризация задачи проведена согласно (1.4) по безразмерной массовой скорости $U_y(t_1,x_1)$ при условии, что $|U_y(t_1,x)|\ll 1$. Это неравенство эквивалентно неравенству $|u_y(t_1,x_1)|\ll v_T$, где $v_T=1/\sqrt{\beta}$ -- тепловая скорость молекул, имеющая порядок скорости звука.

Величину безразмерной массовой скорости $U_y(t_1,x_1)$ согласно ее
определению (1.2):
$$
U_y(t_1,x_1)=\dfrac{1}{\pi^{3/2}}\int \exp(-C^2)C_y\varphi(t_1,x_1,
\mathbf{C})d^3C.
\eqno{(1.5)}
$$

С помощью (1.5) кинетическое линеаризованное уравнение (1.3) записывается в виде:
$$
\dfrac{\partial \varphi}{\partial t_1}+C_x\dfrac{\partial \varphi}{\partial x_1}+\varphi(t_1,x_1,\mathbf{C}) =\dfrac{2C_y}{\pi^{3/2}}
\int\exp(-{C'}^2)C_y'\varphi(t_1, x_1,\mathbf{C'})\,d^3C'.
\eqno{(1.6)}
$$

Сформулируем зеркально--диффузные граничные условия, записанные относительно функции $\varphi(t_1,x_1,\mathbf{C})$:
$$
\varphi(t_1,0,\mathbf{C})=2qC_yU_s(t_1)+(1-q)\varphi(t_1,0,
-C_x,C_y,C_z),\quad C_x>0,
\eqno{(1.7)}
$$
и
$$
\varphi(t_1,x_1\to+\infty,\mathbf{C})=0.
\eqno{(1.8)}
$$

Итак, граничная задача о колебаниях газа сформулирована полностью и состоит в решении уравнения (1.6) с граничными условиями (1.7) и (1.8).

Отметим, что к выражению (1.5) для безразмерной массовой скорости можно придти, исходя из определения размерной массовой скорости газа (1.2). В самом деле, подставляя в (1.2) выражение (1.4), приходим в точности к выражению (1.5).

\begin{center}
\item{}\section{\bf Декомпозиция граничной задачи}
\end{center}

Учитывая, что колебания пластины рассматриваются вдоль оси $y$, будем искать, следуя Черчиньяни \cite{16}, функцию $\varphi(t_1,x_1,\mathbf{C})$ в виде
$$
\varphi(t_1,x_1,\mathbf{C})=C_yH(t_1,x_1,C_x).
\eqno{(2.1)}
$$
Тогда безразмерная массовая скорость (1.5) с помощью (2.1) равна
$$
U_y(t_1,x_1)=\dfrac{1}{2\sqrt{\pi}}\int\limits_{-\infty}^{\infty}
\exp(-C_x'^2)H(t_1,x_1,C_x')dC_x'.
\eqno{(2.2)}
$$

С помощью указанной выше подстановки (2.1) кинетическое уравнение (1.6) преобразуется к виду:
$$
\dfrac{\partial H}{\partial t_1}+C_x\dfrac{\partial H}{\partial x_1}+
H(t_1,x_1,C_x)=\dfrac{1}{\sqrt{\pi}}\int\limits_{-\infty}^{\infty}
\exp(-C_x'^2)H(t_1,x_1,C_x')dC_x'.
\eqno{(2.3)}
$$

Граничные условия (1.7) и (1.8) преобразуются в следующие:
$$
H(t_1,0,C_x)=2qU_s(t_1)+(1-q)H(t_1,0,-C_x),\qquad C_x>0,
\eqno{(2.4)}
$$
$$
H(t_1,x_1\to +\infty, C_x)=0.
\eqno{(2.5)}
$$

Следующим шагом одновременно осуществим комплексификацию кинетического
уравнения и выделим временную переменную, положив далее:
$$
H(t_1,x_1,C_x)=\Re\{e^{-i\omega_1t_1}h(x_1,C_x)\}
\eqno{(2.6)}
$$
и
$$
U_0\cos\omega_1t_1=\Re\{e^{-i\omega_1t_1}U_0\}.
$$

Теперь вместо (2.3) мы получаем комплексно--значное уравнение (уравнение относительно
комплексно--значной функции $h(x_1,C_x)$):
$$
C_x\dfrac{\partial h}{\partial x_1}+(1-i\omega_1)h(x_1,C_x)
=\dfrac{1}{\sqrt{\pi}}\int\limits_{-\infty}^{\infty}
\exp(-C_x'^2)h(x_1,C_x')dC_x'.
\eqno{(2.7)}
$$

Граничные условия (2.4) и (2.5) переходят в следующие:
$$
h(0,C_x)=2qU_0+(1-q)h(0,-C_x),\qquad C_x>0,
\eqno{(2.8)}
$$
и
$$
h(x_1\to+\infty,C_x)=0.
\eqno{(2.9)}
$$

Тогда безразмерная массовая скорость согласно (2.2) и (2.6) равна:
$$
U_y(t_1,x_1)=\dfrac{1}{2\sqrt{\pi}}\int\limits_{-\infty}^{\infty}
\exp(-C_x'^2)\Re\{e^{-i\omega_1t_1}h(x_1,C_x')\}dC_x'.
\eqno{(2.10)}
$$

Мы получили граничную задачу, состоящую в решении уравенния (2.7)
с граничными условиями (2.8) и (2.9). Скорость газа согласно (2.10) будет вычислена во второй части нашей работы.

\begin{center}
\item{}\section{ \bf Собственные решения непрерывного спектра}
\end{center}

Перепишем граничную задачу (2.7), (2.8) и (2.9) в виде:
$$
\mu\dfrac{\partial h}{\partial x_1}+z_0h(x_1,\mu)=\dfrac{1}{\sqrt{\pi}}
\int\limits_{-\infty}^{\infty}\exp(-{\mu'}^2)h(x_1,\mu')d\mu',
\eqno{(3.1)}
$$
где
$$
z_0=1-i\omega_1,
$$
и
$$
h(0,\mu)=2qU_0+(1-q)h(0,-\mu)d\mu,\qquad \mu>0,
\eqno{(3.2)}
$$
$$
h(+\infty,\mu)=0.
\eqno{(3.3)}
$$

Граничная задача (3.1)--(3.3) будет рассмотрена в следующей части нашей работы.
Разделение переменных в уравнении (3.1) осуществляется следующей подстановкой
$$
h_\eta(x_1,\mu)=\exp\Big(-\dfrac{x_1z_0}{\eta}\Big)\Phi(\eta,\mu),
\eqno{(3.4)}
$$
где $\eta$ -- параметр разделения, или спектральный параметр, вообще говоря, комплексный.

Подставляя (3.4) в уравнение (3.1) получаем характеристическое уравнение
$$
(\eta-\mu)\Phi(\eta,\mu)=\dfrac{\eta}{\sqrt{\pi}z_0}
\int\limits_{-\infty}^{\infty}
\exp(-{\mu'}^2)\Phi(\eta,\mu')d\mu'.
\eqno{(3.5)}
$$
Если ввести обозначение
$$
n(\eta)=\dfrac{1}{z_0}\int\limits_{-\infty}^{\infty}
\exp(-{\mu'}^2)\Phi(\eta,\mu')d\mu',
\eqno{(3.6)}
$$
то уравнение (3.5) может быть записано с помощью (3.6) в виде
$$
(\eta-\mu)\Phi(\eta,\mu)=\dfrac{1}{\sqrt{\pi}}\eta n(\eta),\qquad
\eta\in \mathbb{C}.
\eqno{(3.7)}
$$
Уравнение (3.7) является конечным (недифференциальным) уравнением.
Условие (3.6) называется нормировочным условием,
нормировочным интегралом, или просто нормировкой.

Решение характеристического уравнения для действительных значений
параметра $\eta$ будем искать в пространстве
обобщенных функций \cite{6}.
Обобщенное решение уравнения (3.7) имеет вид:
$$
\Phi(\eta,\mu)=\dfrac{1}{\sqrt{\pi}}\eta n(\eta)P\dfrac{1}{\eta-\mu}+
g(\eta)\delta(\eta-\mu),
\eqno{(3.8)}
$$
где $-\infty<\eta, \mu <+\infty$.

Здесь $g(\eta)$ -- произвольная непрерывная функция, определяемая из
условия нормировки, $\delta(x)$ -- дельта--функция Дирака, символ $Px^{-1}$
означает главное значение интеграла при интегрировании $x^{-1}$.
Подставляя (3.8) в (3.6), получаем уравнение, из которого находим
$$
n(\eta)\lambda(\eta)=\exp(-\eta^2)g(\eta),
$$
где $\lambda(z)$ -- дисперсионная функция, введенная равенством
$$
\lambda(z)=1-i\omega_1+\dfrac{z}{\sqrt{\pi}}\int\limits_{-\infty}^{\infty}
\dfrac{\exp(-\tau^2)d\tau}{\tau-z}.
$$
Эту функцию можно преобразовать к виду: $\lambda(z)=-i\omega_1+\lambda_0(z)$,
где $\lambda_0(z)$ -- известная функция из теории плазмы,
$$
\lambda_0(z)=\dfrac{1}{\sqrt{\pi}}\int\limits_{-\infty}^{\infty}
\dfrac{e^{-\tau^2}\tau d\tau}{\tau-z}.
$$
Собственные функции (3.8) определены с точностью
до мультипликативной "постоянной"\,$n(\eta)$:
$$
\Phi(\eta,\mu)=\Big[\dfrac{1}{\sqrt{\pi}}\eta P\dfrac{1}{\eta-\mu}+
\exp(\eta^2)\lambda(\eta)\delta(\eta-\mu)\Big]n(\eta).
\eqno{(3.9)}
$$

Собственные функции (3.9) называются собственными функциями непрерывного
спектра, ибо спектральный параметр $\eta$ непрерывным образом заполняет всю действительную прямую.

Далее в силу однородности уравнения (3.1) можно считать, что
$$
n(\eta)\equiv 1.
$$

Таким образом, собственные решения уравнения (3.4) имеют вид
$$
h_\eta(x,\mu)=\exp\Big(-\dfrac{x_1}{\eta}z_0\Big)
\Big[\dfrac{1}{\sqrt{\pi}}\eta P\dfrac{1}{\eta-\mu}+
\exp(\eta^2)\lambda(\eta)\delta(\eta-\mu)\Big].
\eqno{(3.10)}
$$

Собственные решения (3.10) отвечают непрерывному спектру характеристического
уравнения, ибо спектральный параметр непрерывным образом пробегает всю числовую прямую, т.е. непрерывный спектр $\sigma_c$
есть вся конечная часть числовой прямой: $\sigma_c=(-\infty,+\infty)$.

По условию задачи мы ищем решение, невозрастающее вдали от стенки.
Поэтому далее будем рассматривать положительную часть непрерывного спектра. В этом случае собственные решения (3.10) являются исчезающими вдали от стенки. В связи с этим спектром граничной задачи будем называть положительную действительную полуось параметра $\eta$:
$\sigma_c^{\rm problem}=(0,+\infty)$.

В заключение этого п. приведем формулы Сохоцкого для дисперсионной функции:
$$
\lambda^{\pm}(\mu)=\pm i\sqrt{\pi}\mu e^{-\mu^2}-i\omega_1+
\dfrac{1}{\sqrt{\pi}}\int\limits_{0}^{\infty}
\dfrac{e^{-\tau^2}\tau d\tau}{\tau-\mu}.
$$
Разность граничных значений дисперсионной функции отсюда равна:
$$
\lambda^+(\mu)-\lambda^-(\mu)=2\sqrt{\pi}\mu e^{-\mu^2}i,
$$
полусумма граничных значений равна:
$$
\dfrac{\lambda^+(\mu)+\lambda^-(\mu)}{2}=-i\omega_1+\dfrac{1}{\sqrt{\pi}}
\int\limits_{0}^{\infty}\dfrac{e^{-\tau^2}\tau d\tau}{\tau-\mu}.
$$

Заметим, что на действительной оси действительная часть
дисперсионной функции $\lambda_0(\mu)$ имеет два нуля $\pm\mu_0$, $\mu_0=0.924\cdots$. Эти два нуля в силу четности функции $\lambda_0(\mu)$ различаются лишь знаками.

Отметим, что на действительной оси дисперсионную функцию удобнее использовать в численных расчетах в виде (см. \cite{19})
$$
\lambda_0(\mu)=1-2\mu^2 \int\limits_{0}^{1}\exp(-\mu^2(1-t^2))dt,\qquad
\mu\in(-\infty,+\infty).
$$

\begin{center}
\item{}\section{\bf Собственные решения дискретного спектра}
\end{center}

Разложим дисперсионную функцию в ряд Лорана по отрицательным степеням
переменного $z$ в окрестности бесконечно удаленной точки:
$$
\lambda(z)=-i\omega_1-\dfrac{1}{2z^2}-\dfrac{3}{4z^4}-\dfrac{15}{8z^6}-\cdots,
\quad z\to \infty.
\eqno{(4.1)}
$$

Из разложения (4.1) видно, что при малых значениях $\omega_1$
дисперсионная функция имеет два отличающиеся лишь знаками комплексно--значных нуля:
$$
\pm\eta_0^{(0)}(\omega_1)=\dfrac{1+i}{2\sqrt{\omega_1}}.
$$

\begin{figure}[t]
\begin{center}
\includegraphics[width=17.0cm, height=12cm]{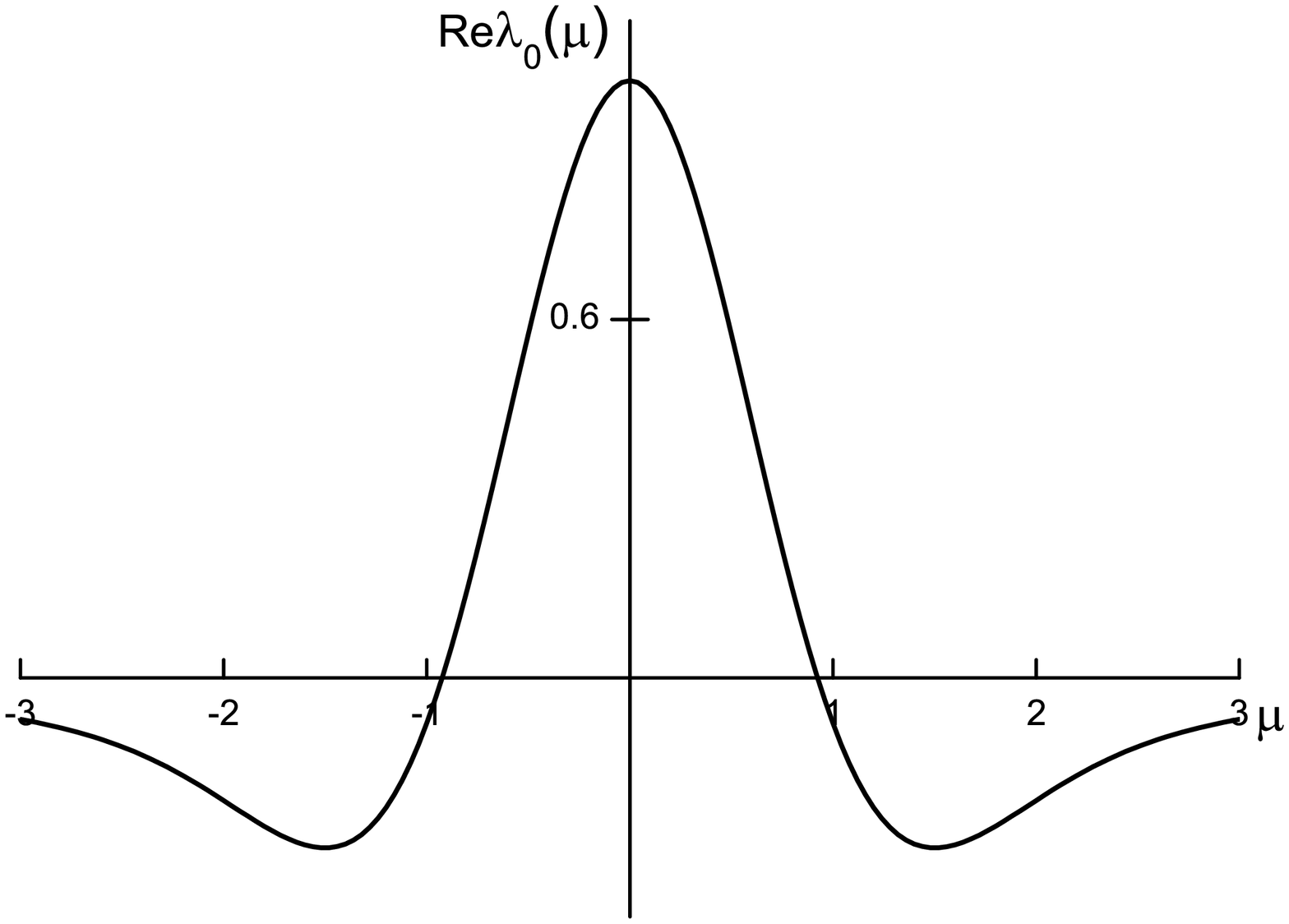}
\end{center}
\begin{center}
{{Рис 1. Дисперсионная функция $\lambda_0(\mu)$ на действительной оси.}}
\end{center}
\end{figure}

\begin{figure}[h]
\begin{center}
\includegraphics[width=17.0cm, height=12cm]{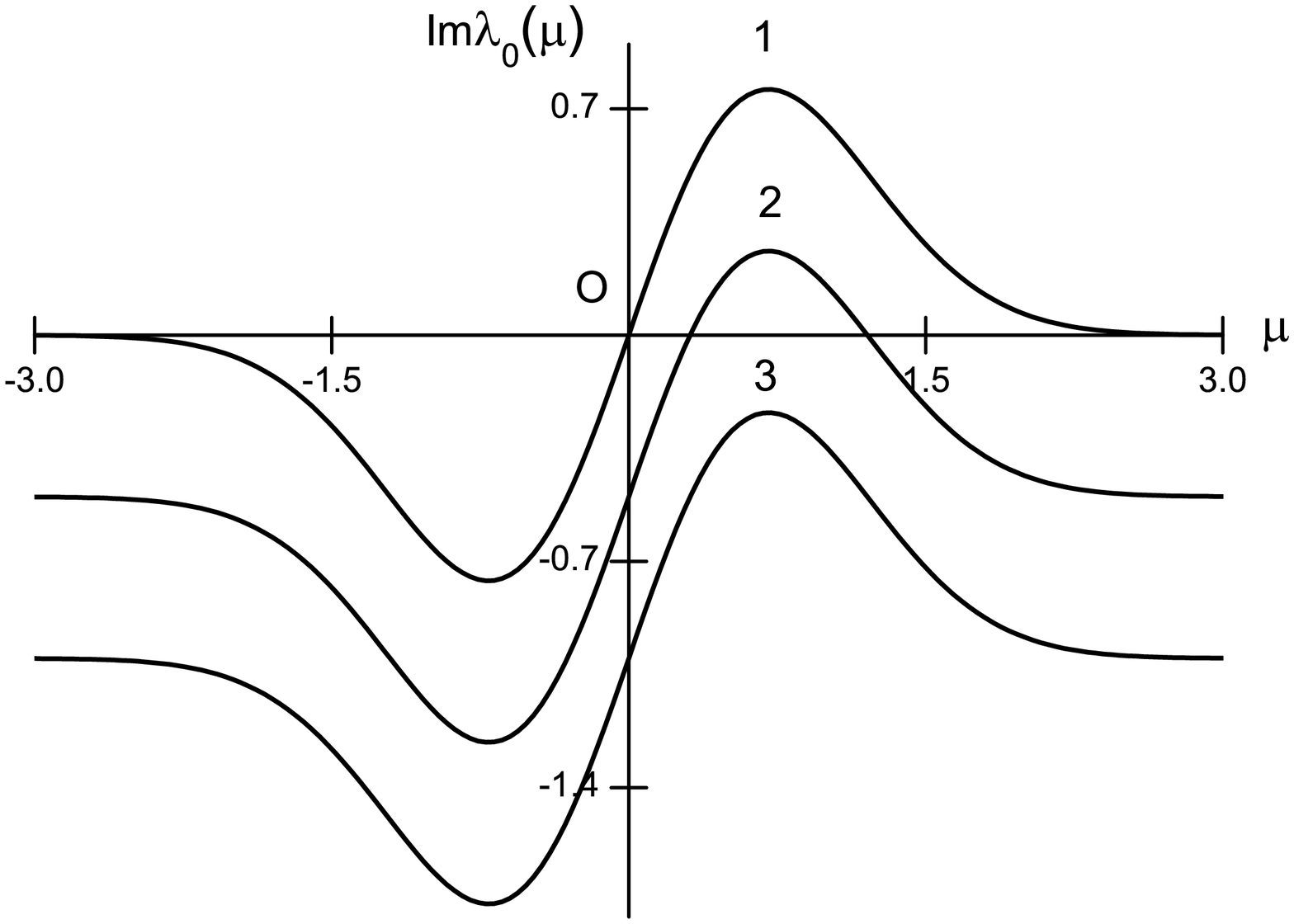}
\end{center}
\begin{center}
{{Рис 1. Мнимая часть дисперсионной функции $\lambda^+(\mu)$ на действительной оси. Кривые $1,2,3$ отвечают значениям параметра $\omega_1=0,0.5,1$.}}
\end{center}
\end{figure}
Отсюда видно, что при $\omega_1\to 0$ оба нуля дисперсионной функции
имеют пределом одну бесконечно удаленную точку $\eta_i=\infty$ кратности (порядка) два.

Из разложения (4.1) видно так же, что значение дисперсионной функции в
бесконечно удаленной точки равно:
$$
\lambda(\infty)=-i\omega_1.
$$

Применим теперь принцип аргумента для нахождения нулей дисперсионной
функции в верхней и нижней полуплоскостях. Этот подход является наиболее
общим.

Возьмем две прямые $\Gamma_\varepsilon^{\pm}$, параллельные
действительной оси и отстоящие от нее на расстоянии $\varepsilon,\;
\varepsilon>0$. Число $\varepsilon$ выберем настолько малым, чтобы
все нули дисперсионной функции лежали вне узкой полосы, заключенной
между прямыми $\Gamma_\varepsilon^{+}$ и $\Gamma_\varepsilon^{-}$.

Согласно принципу аргумента разность между числом нулей и числом
полюсов дисперсионной функции равно приращению ее логарифма:
$$
N-P=\dfrac{1}{2\pi i}\Bigg[\;\int\limits_{\Gamma_\varepsilon^+}+
\int\limits_{\Gamma_\varepsilon^-}\Bigg]\,d\,\ln \lambda(z).
\eqno{(4.2)}
$$

В (4.2) каждый нуль и полюс считаются столько раз, какова их
кратность, прямые $\Gamma_\varepsilon^{+}$ и
$\Gamma_\varepsilon^{-}$ проходятся соответственно в положительном и
отрицательном напрвлениях. Ясно, что дисперсионная функция полюсов не
имеет, т. е. $P=0$.

В пределе при $\varepsilon\to 0$ из равенства (4.2) получаем:
$$
N=\dfrac{1}{2\pi i}\int\limits_{-\infty}^{\infty}\,
d\ln \dfrac{\lambda^+(\mu)}{\lambda^-(\mu)}.
\eqno{(4.3)}
$$

Интеграл из (4.3) разобьем на два:
$$
\int\limits_{-\infty}^{\infty}
d\ln \dfrac{\lambda^+(\mu)}{\lambda^-(\mu)}=
\int\limits_{0}^{\infty}
d\ln \dfrac{\lambda^+(\mu)}{\lambda^-(\mu)}+
\int\limits_{-\infty}^{0}
d\ln \dfrac{\lambda^+(\mu)}{\lambda^-(\mu)}.
$$

Во втором интеграле сделаем замену переменной: $\tau\to -\tau$, при
такой замене имеем:
$$
\lambda^+(-\tau)=\lambda^-(\tau), \qquad
\lambda^-(-\tau)=\lambda^+(\tau).
$$
Следовательно, второй интеграл равен первому. В самом деле,
$$
\int\limits_{-\infty}^{0}
d\ln \dfrac{\lambda^+(\mu)}{\lambda^-(\mu)}=-
\int\limits_{0}^{\infty}
d\ln \dfrac{\lambda^+(-\mu)}{\lambda^-(-\mu)}=
-\int\limits_{0}^{\infty}
d\ln \dfrac{\lambda^-(\mu)}{\lambda^+(\mu)}=
\int\limits_{0}^{\infty}
d\ln \dfrac{\lambda^+(\mu)}{\lambda^-(\mu)}.
$$
Таким образом,
$$
N=\dfrac{1}{\pi i}\int\limits_{0}^{\infty}\,d\ln
\dfrac{\lambda^+(\mu)}{\lambda^-(\mu)}.
\eqno{(4.4)}
$$

Рассмотрим теперь на комплексной плоскости семейство кривых $\Gamma=\Gamma(\omega_1):
\; z=G(t), \;0\leqslant t \leqslant +\infty$, где
$$
G(t)=\dfrac{\lambda^+(t)}{\lambda^-(t)}.
$$
Нетрудно проверить, что
$$
G(0)=1,\qquad \lim\limits_{\tau\to +\infty} G(t)=1.
$$

Эти равенства означают, что кривые $\Gamma(\omega_1)$ являются замкнутыми: они выходят из точки $z=1$ и заканчиваются в этой точке.
Согласно (4.4) имеем:
$$
N=\dfrac{1}{\pi i}\Big[\ln |G(\tau)|+i\arg G(\tau)\Big]
_0^{+\infty}=\dfrac{1}{\pi}\Big[\arg G(\tau)
\Big]_0^{+\infty}.
$$
Учитывая предыдущие равенства, отсюда получаем:
$$
N=\dfrac{1}{\pi}\Big[\arg G(t)\Big]_0^{+\infty}=2\varkappa(G),
\eqno{(4.5)}
$$
или
$$
N=2\varkappa(G),
$$
где $\varkappa=\varkappa(G)$ -- индекс функции $G(t)$ -- число
оборотов кривой $\Gamma(\omega_1)$ относительно начала координат, совершаемых
в положительном направлении.

Из формулы (4.5) видно, что
$$
N=\dfrac{1}{\pi}\Big[\arg G(+\infty)-\arg G(0)\Big]=
\dfrac{1}{\pi}\arg G(+\infty),
\eqno{(4.6)}
$$
ибо $\arg G(0)=0$.

\begin{figure}[t]
\begin{center}
\includegraphics[width=17.0cm, height=12cm]{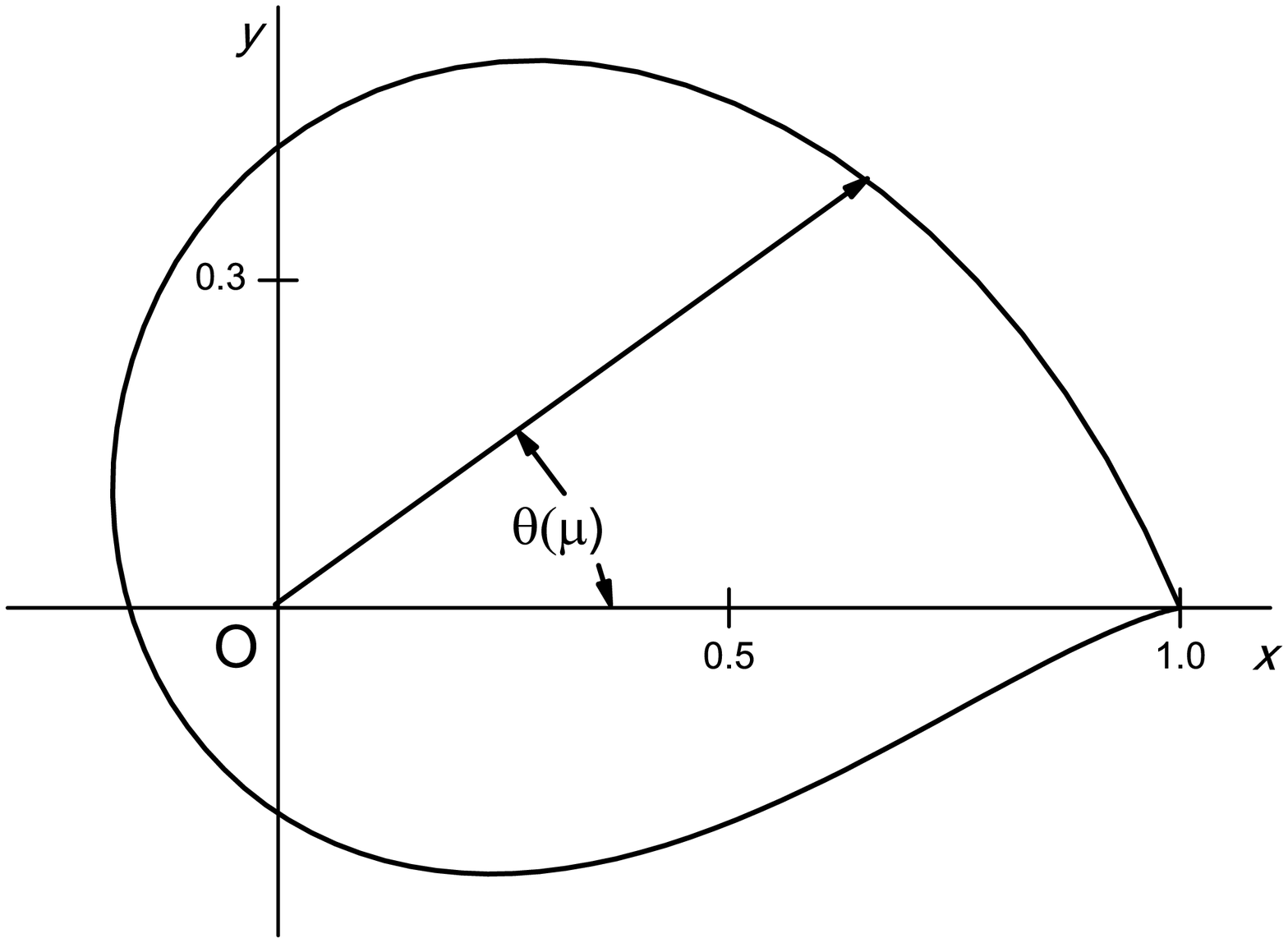}
\end{center}
\begin{center}
{{Рис 3. Кривая $\Gamma(\omega_1)$ является замкнутой и охватывает начало координат при $0<\omega_1<\omega_1^*$. Индекс функции $G(\mu)$ равен единице, дисперсионная функция имеет два комплексно--значных нуля.}}
\end{center}
\end{figure}

Введем угол $ \theta( \mu)= \arg G(\mu)$ --
главное значение аргумента функции $G(\mu)$, фиксированное
в нуле условием $ \theta(0)=0$.

Обозначим: $s(\mu)=\sqrt{\pi}\mu e^{-\mu^2}$.
Выделим действительную и мнимую части функции $G(t)$:
$$
G(t)=\dfrac{\lambda_0(\mu)-i\omega_1+is(\mu)}{\lambda_0(\mu)-i\omega_1+is(\mu)}=
$$
$$+
\dfrac{\lambda_0^2(\mu)-s^2(\mu)+\omega_1^2}{\lambda_0^2(\mu)+[\omega_1+
s(\mu)]^2}+i\dfrac{2\lambda_0(\mu)s(\mu)}{\lambda_0^2(\mu)+[\omega_1+
s(\mu)]^2}.
$$

\begin{figure}[t]
\begin{center}
\includegraphics[width=17.0cm, height=12cm]{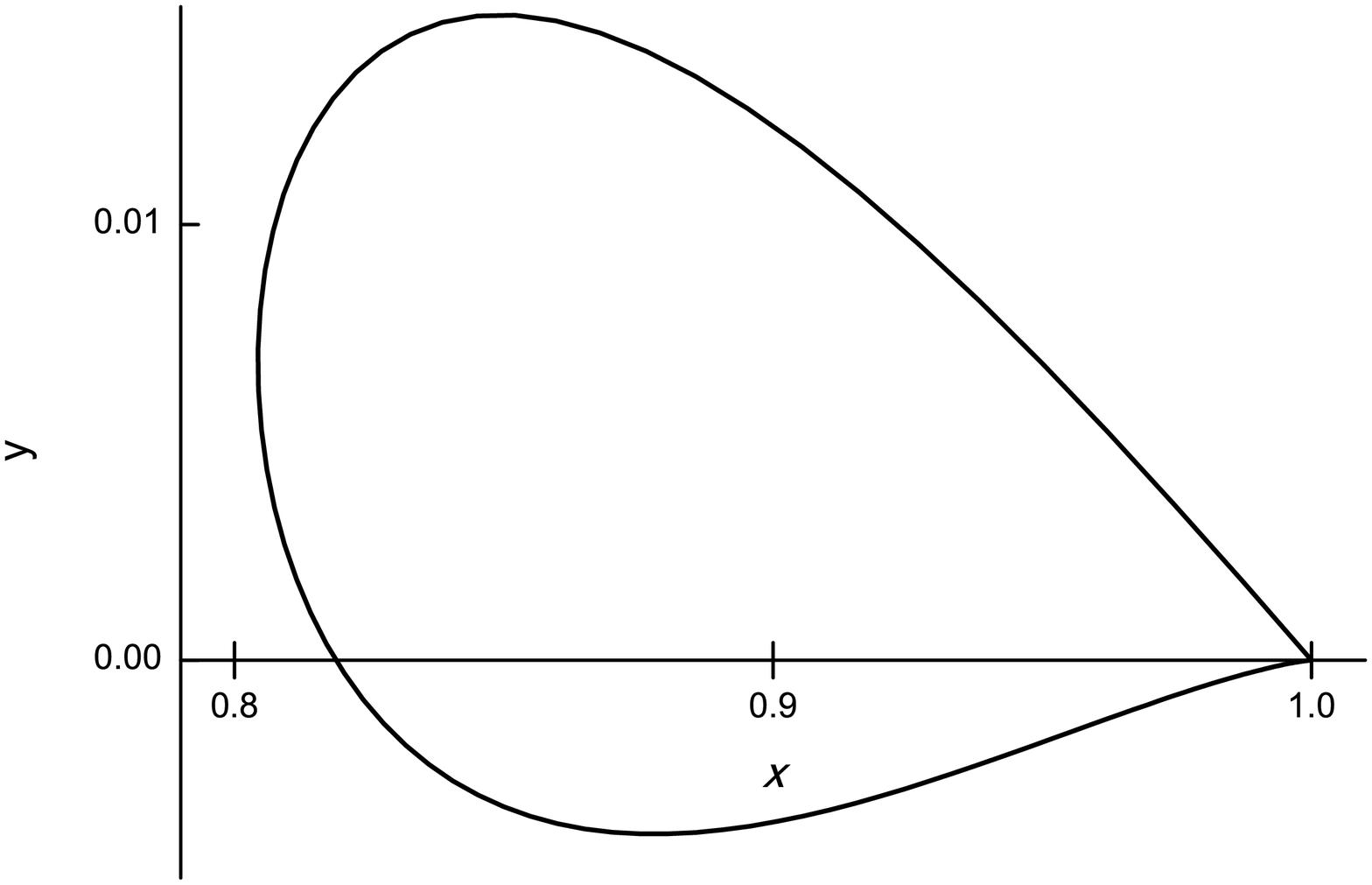}
\end{center}
\begin{center}
{{Рис 4. Кривая $\Gamma(\omega_1)$ не охватывает начало координат при $\omega_1>\omega_1^*$. Индекс функции $G(\mu)$ равен нулю, дисперсионная функция имеет не имеет нулей в верхней и нижней полуплоскостях.}}
\end{center}
\end{figure}

Отсюда видно, что
$$
\Re G(\mu)=\dfrac{\lambda_0^2(\mu)-s^2(\mu)+\omega_1^2}{\lambda_0^2(\mu)+[\omega_1+
s(\mu)]^2}, \quad
\Im G(\mu)=\dfrac{2\lambda_0(\mu)s(\mu)}{\lambda_0^2(\mu)+[\omega_1+
s(\mu)]^2}.
\eqno{(4.7)}
$$

Введем выделенную частоту колебаний пластины, ограничивающей газ:
$$
\omega_1^*=\max\limits_{0\leqslant \mu<+\infty}\sqrt{-\lambda_0^2(\mu)+s^2(\mu)}\approx 0.733.
$$

Эту частоту колебаний будем называть {\it критической}.

Покажем, что в случае, когда частота колебаний пластины меньше критической, т.е. при $0\leqslant \omega <\omega_1^*$, индекс функции $G(t)$ равен единице. Это означает, что согласно (4.6) число комплексно--значных нулей дисперсионной функции в разрезанной комплексной плоскости с разрезом вдоль действительной оси, равно двум.

В случае, когда частота колебаний пластины превышает критическую ($\omega>\omega_1^*$) индекс функции $G(t)$ равен нулю: $\varkappa(G)=0$. Это означает, что дисперсионная функция не имеет нулей в верхней и нижней полуплоскостях. В этом случае дискретных (частных) решений исходное кинетическое уравнение (3.1) не имеет.

\begin{figure}[h]
\begin{center}
\includegraphics[width=17.0cm, height=12cm]{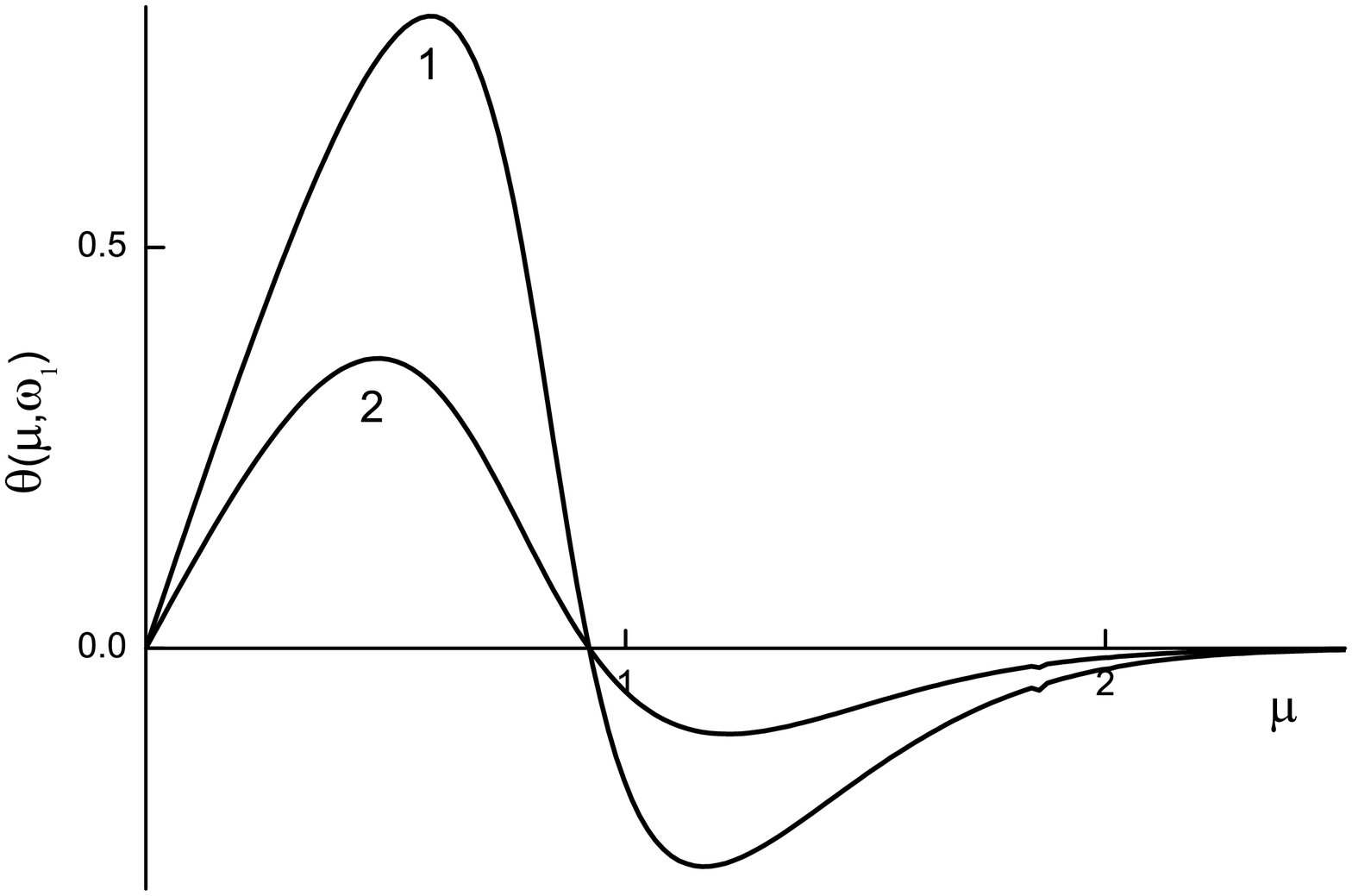}
\end{center}
\begin{center}
{{Рис 5. Зависимость угла $\theta=\theta(\mu,\omega_1)$ от $\mu$ при различных значениях параметра $\omega_1$ при $\omega_1>\omega_1^*$. Индекс функции $G(\mu)$ равен нулю. Приращение угла на полуоси равно нулю. Кривые $1$ и $2$ отвечает значениям $\omega_1=1$ и $\omega_1=1.5.$}}
\end{center}
\end{figure}

Кривые $\Gamma(\omega_1)$ согласно (4.7) определяются параметрическими уравнениями
$$
\Gamma(\omega_1): \quad x=\Re G(\mu),\quad y=\Im G(\mu),\quad
0 \leqslant \mu \leqslant +\infty.
\eqno{(4.8)}
$$
При $\omega_1=0$ мы имеем случай, рассмотренный в \cite{19}. В этом случае кривая $\Gamma(0)$ охватывает один раз начало координат. В самом деле, функция $\lambda_0(\mu)$ имеет единственный нуль $\mu_0\approx 0.924$ на действительной оси, причем $\lambda_0(\mu)>0$ при $0\leqslant \mu<\mu_0$ и $\lambda_0(\mu)<0$ при $\mu_0<\mu<+\infty$. Функция $y_1(\mu)=\lambda_0^2(\mu)-s^2(\mu)$ имеет два нуля $\mu_1\approx 0.447$ и $\mu_2\approx 1.493$. При этом $y_1(\mu)>0$ при $\mu\in [0,\mu_1)\cup (\mu_1,+\infty)$, а при $\mu\in (\mu_1,\mu_2)$: $y_1(\mu)<0$.

\begin{figure}[h]
\begin{center}
\includegraphics[width=17.0cm, height=12cm]{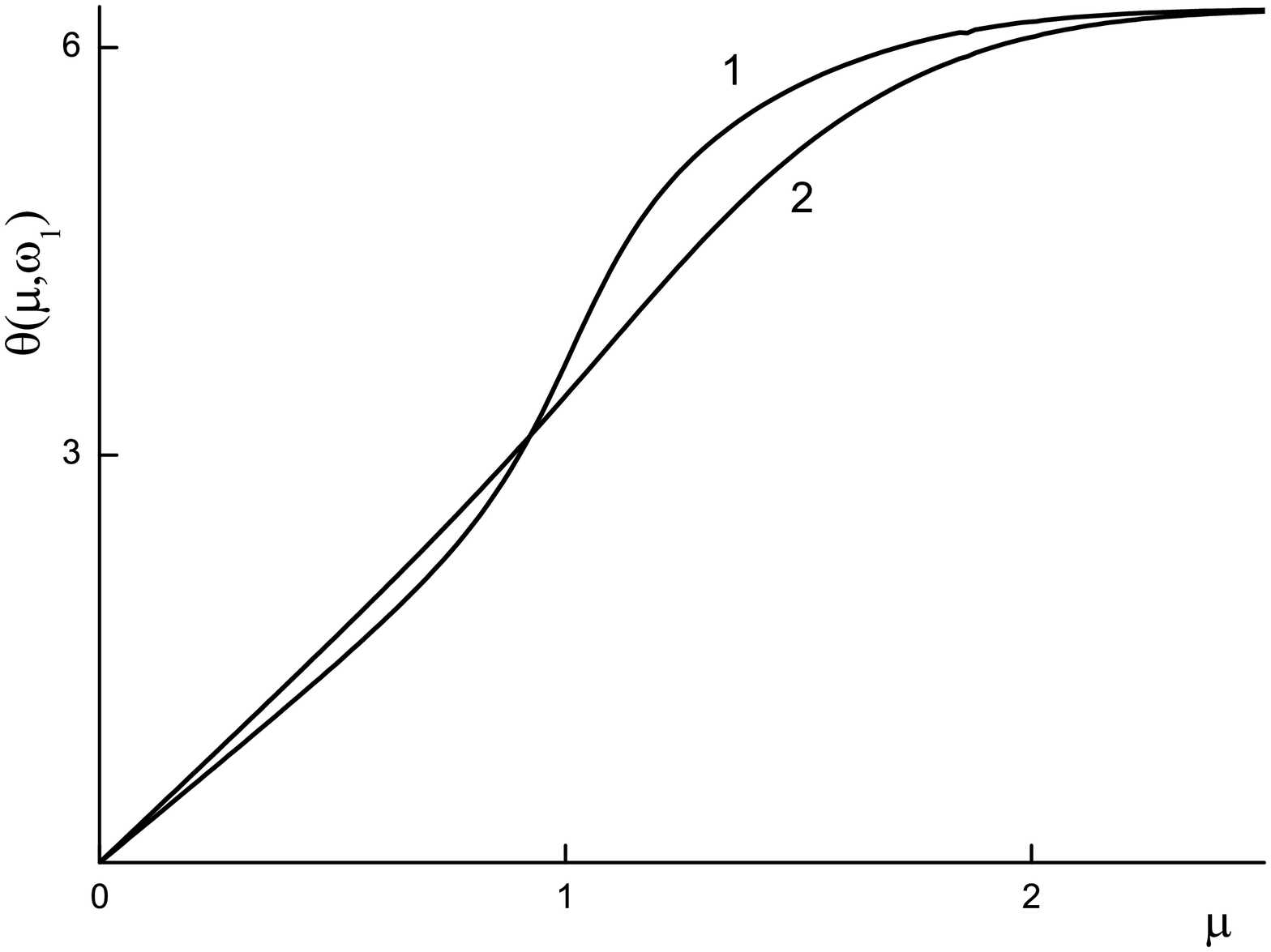}
\end{center}
\begin{center}
{{Рис 6. Зависимость угла $\theta=\theta(\mu,\omega_1)$ от $\mu$ при различных значениях параметра $\omega_1$ при $0<\omega_1<\omega_1^*$. Индекс функции $G(\mu)$ равен единице. Приращение угла на полуоси равно $2\pi$. Кривые $1$ и $2$ отвечает значениям $\omega_1=0.5$ и $\omega_1=0.3.$}}
\end{center}
\end{figure}

Теперь из соотношений (4.7) и (4.8) видно, что при изменении $\mu$ от $0$ до $\mu_1$ кривая $\Gamma(0)$ выходит из точки $z=1$ и при $\mu=\mu_1$ оказывается в точке на мнимой оси с координатой
$$
y(\mu_1)=\Im G(\mu_1)=\dfrac{2\lambda_0(\mu_1)s(\mu_1)}{\lambda_0^2(\mu_1)+s^2(\mu_1)}>0.
$$
При этом кривая $\Gamma(0)$ описывает дугу, лежащую в первой четверти. При изменении $\mu$ от $\mu_1$ до $\mu_0$ кривая описывает дугу, лежащую во второй четверти, и при $\mu=\mu_0$ оказывается в точке
на действительной оси с координатой $x(\mu_0)=-1$. При измененении $\mu$ от $\mu_0$ до $\mu_2$ кривая $\Gamma(0)$ описывает дугу, лежащую в третьей четверти и при $\mu=\mu_2$ оказывается в точке на мнимой оси с координатой
$$
y(\mu_2)=\dfrac{2\lambda_0(\mu_2)s(\mu_2)}{\lambda_0^2(\mu_2)+s^2(\mu_2)}<0,
\quad y(\mu_2)\geqslant -1.
$$

При дальнейшем изменении $\mu$ от $\mu_2$ до $+\infty$ кривая $\Gamma(0)$
лежит в четвертой четверти и заканчивается в точке $z=1$, описывая один оборот вокруг начала координат.

Пусть теперь параметр $\omega_1$ изменяется в пределах от нуля до значения $\omega_1^\circ=s(\mu_0)=\sqrt{\pi}\mu_0e^{-\mu_0^2}\approx 0.697$. Теперь корни $\mu_1$ и $\mu_2$ уравенения
$$
y_1(\mu,\omega_1)=\lambda_0^2(\mu)-s^2(\mu)+\omega_1^2
$$
становятся функциями параметра $\omega_1$: $\mu_1=\mu_1(\omega_1)$ и
$\mu_2=\mu_2(\omega_1)$, причем $\mu_1^2(\omega_1)< \mu_2(\omega_1)$.
Нетрудно понять, что семейство кривых $\Gamma(\omega_1)$ охватывает начало координат тогда и только тогда, когда для нулей $\mu_1(\omega_1), \mu_2$ и $\mu_2(\omega_1)$ выполняется неравенство
$$
\mu_1(\omega_1)<\mu_2<\mu_2(\omega_1).
$$
Точка $\mu_2(\omega_1)$ не зависит от $\omega_1$, а при возрастании $\omega_1$ от $0$ до $\omega_1^\circ$ точки $\mu_1(\omega_1)$ и $\mu_2(\omega_1)$ сближаются навстречу друг другу. При $\omega_1=\omega_1^\circ$ точки $\mu_0$ и $\mu_2(\omega_1)$ совпадают. Это означает, что кривая $\Gamma(\omega_1^\circ)$ проходит через начало координат. Этому случаю можно приписать индекс $\Gamma(\mu_0)=1/2$.
\begin{figure}[h]
\begin{center}
\includegraphics[width=17.0cm, height=12cm]{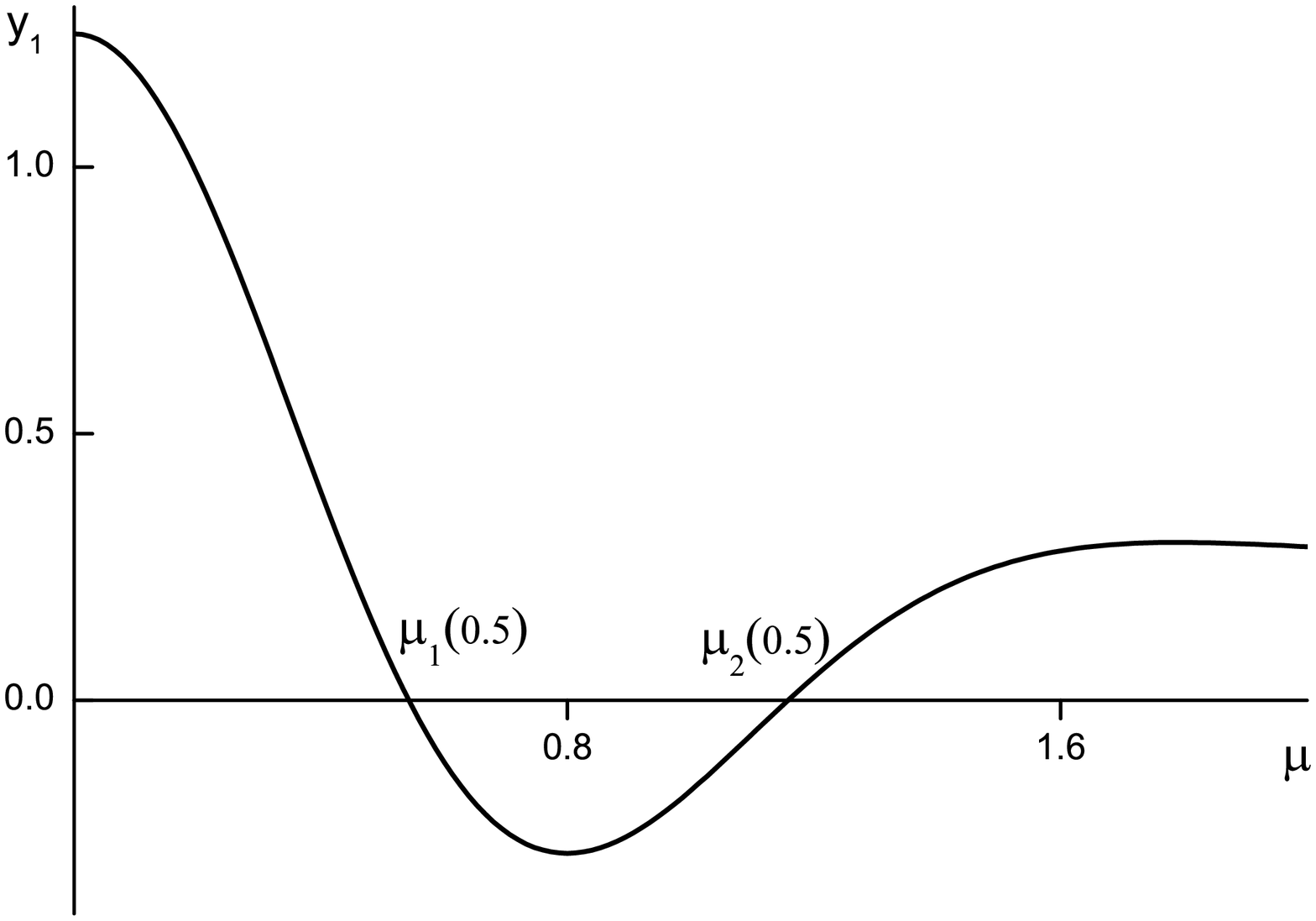}
\end{center}
\begin{center}
{{Рис 7. Нули $\mu_1(\omega_1)$ и $\mu_2(\omega_1)$ функции $y_1(\mu,\omega_1)$ при $\omega_1=0.5$; $\mu_1(0.5)=0.543,\quad \mu_2(0.5)=1.158$.}}
\end{center}
\end{figure}

При дальнейшем возрастании $\omega_1$ от $\omega_1^\circ$ до $\omega_1^*$ точка $\mu_2(\omega_1)$ сначала перемещается влево от точки $\mu_0$, не зависящей от $\omega_1$, а далее точки $\mu_1(\omega_1)$ и
$\mu_2(\omega_1)$ совпадают:
$$
\mu_1(0.733)=\mu_2(0.733)=0.799.
$$
Затем при $\omega \geqslant \omega_1^*$ функция $x(\mu)$ становится положительной при всех значениях $\mu, \;\mu \geqslant 0$.
Итак, при $0\leqslant \omega_1<\omega_1^*$ индекс равен единице: $\varkappa(G)=1$, а при $\omega>\omega_1^*$ индекс равен нулю: $\varkappa(G)=0$. Это означает, что при $0 \leqslant \omega_1<\omega_1^*$ дисперсионная функция имеет два нуля, а при $\omega_1>\omega_1^*$ дисперсионная функция нулей в верхней и нижней помлексной полуплоскости не имеет.

При $0 \leqslant \omega_1<\omega_1^*$ нули дисперсионной функции обозначим через $\eta_0(\omega_1)$ и $-\eta_0(\omega_1)$. В силу четности дисперсионной функции ее конечные нули различаются только знаками, имея одинаковые модули.

Таким образом, дискретный спектр характеристического уравнения, состоящий из нулей дисперсионной функции, в случае $0\leqslant \omega_1<\omega_1^*$ есть множество из двух точек $\sigma_d(\omega_1)=\{\eta_0(\omega_1),
-\eta_0(\omega_1)\}$. При $\omega_1>\omega_1^*$ дискретный спектр --- это пустое множество. При $0\leqslant \omega_1<\omega_1^*$ собственными функциями характеристического уравнения являются следующие два решения характеристического уравнения:
$$
\Phi(\pm \eta_0(\omega_1),\mu)=\dfrac{1}{\sqrt{\pi}}\dfrac{\pm \eta_0(\omega_1)}{\pm \eta_0(\omega_1)-\mu}
$$
и два соответствующих собственных решения исходного характеристического уравнения (3.1):
$$
h_{\pm \eta_0(\omega_1)}(x_1,\mu)=\exp \Big(-\dfrac{x_1z_0}{\pm \eta_0(\omega_1)}\Big)\dfrac{1}{\sqrt{\pi}}\dfrac{\pm \eta_0(\omega_1)}{\pm \eta_0(\omega_1)-\mu}.
$$

Под $\eta_0(\omega_1)$ будем понимать тот из нулей дисперсионной функции, который обладает свойством:
$$
\Re \dfrac{1-i\omega_1}{\eta_0(\omega_1)}>0.
$$
Для этого нуля убывающее собственное решение кинетического уравнения (3.1) имеет вид
$$
h_{\eta_0(\omega_1)}(x_1,\mu)=\dfrac{1}{\sqrt{\pi}}
\exp\Big(-\dfrac{x_1z_0}{\eta_0(\omega_1)}\Big)\dfrac{\eta_0(\omega_1)}
{\eta_0(\omega_1)-\mu}.
$$

Это означает, что дискретный спектр рассматриваемой граничной задачи состоит из одной точки $\sigma_d^{\rm problem}=\{\eta_0(\omega_1)\}$ в случае $0 <\omega_1<\omega_1^*$.

При $\omega_1\to 0$ оба нуля, как уже указывалось выше, перемещаются в одну и ту же бесконечно удаленную точку. Это значит, что в этом случае дискретный спектр характеристического уравнения состоит из одной бесконечно удаленной точки кратности два:
$\sigma_d(0)=\eta_i=\infty$ и является присоединенным к непрерывному спектру. Этот спектр является также и спектром рассматриваемой граничной задачи. Однако, в этом случае дискретных (частных) решения ровно два:
$$
h_1(x_1,\mu)=1, \qquad h_2(x_1,\mu)=x_1-\mu.
$$

Составим общее решение уравнения (3.1) в виде суммы частного (дискретного) решения, убывающего вдали от стенки, и интеграла по
непрерывному спектру от собственных решений, отвечающих непрерывному спектру:
$$
h(x_1,\mu)=\dfrac{a_0}{\eta_0-\mu}\exp\Big(-\dfrac{x_1z_0}
{\eta_0}\Big)
+\int\limits_{0}^{\infty}
\exp\Big(-\dfrac{x_1z_0}{\eta}\Big)\Phi(\eta,\mu)a(\eta)d\eta.
\eqno{(4.9)}
$$
Здесь $a_0$ -- неизвестный постоянный коэффициент, называемый коэффициентом дискретного спектра, $a(\eta)$ -- неизвестная функция, называемая коэффициентом непрерывного спектра, $\Phi(\eta,\mu)$ -- собственные функции характеристического уравнения,
отвечающие непрерывному спектру и единичной нормировке.

Разложение (4.9) можно представить в явном виде:
$$
h(x_1,\mu)=\dfrac{a_0}{\eta_0-\mu}\exp\Big(-\dfrac{x_1z_0}
{\eta_0}\Big)+
$$
$$
+\int\limits_{0}^{\infty}
\exp\Big(-\dfrac{x_1}{\eta}z_0\Big)
\Big[\dfrac{1}{\sqrt{\pi}}\eta P\dfrac{1}{\eta-\mu}+
\exp(\eta^2)\lambda(\eta)\delta(\eta-\mu)\Big]a(\eta)d\eta.
\eqno{(4.10)}
$$
Функция $a(\eta)$ подлежит нахождению из граничных условий (3.2) и (3.3).

Разложение (4.10) можно представить в классическом виде:
$$
h(x_1,\mu)=\dfrac{a_0}{\eta_0-\mu}\exp\Big(-\dfrac{x_1z_0}
{\eta_0}\Big)+
$$
$$
+\dfrac{1}{\sqrt{\pi}}\int\limits_{0}^{\infty}
\exp\Big(-\dfrac{x_1z_0}{\eta}\Big)\dfrac{\eta a(\eta)d\eta}{\eta-\mu}+
\exp\Big(-\dfrac{x_1z_0}{\mu}+\mu^2\Big)\lambda(\mu)a(\mu)\theta_+(\mu),
\eqno{(4.11)}
$$
где $\theta_+(\mu)$ -- функция Хэвисайда,
$$
\theta_+(\mu)=\left\{\begin{array}{c}
                       1,\qquad \mu>0, \\
                       0,\qquad \mu<0
                     \end{array}.\right.
$$

\begin{center}
\item{}\section*{\bf Заключение} 
\end{center}

В настоящей работе сформулирована вторая задача Стокса --- задача о поведении разреженного газа, занимающего полупространство над стенкой, совершающей гармонические колебания. Рассматриваются диффузные граничные условия.
Используется линеаризованное кинетическое уравнение, полученное в результате линеаризации модельного кинетического уравнения Больцмана в
релаксационном приближении.
Отыскивается структура дискретного и непрерывного спектров задачи. Находятся собственные решения, отвечающие дискретному и непрерывному спектрам. Составляется общее решение кинетического уравнения в виде разложения по собственным решениям.

\newpage
\makeatother {\renewcommand{\baselinestretch}{1.2}

 \end{document}